\documentclass[pra,12pt,tightenlines,superscriptaddress,secnumarabic]{revtex4}

\usepackage{bm,amssymb,amsmath,amsthm,mathrsfs,multirow,longtable,url,colonequals,verbatim,afterpage,ifmtarg,pgffor}

\usepackage[continuous]{pagenote}
\makepagenote

\usepackage{hyperref}
\makeatletter
\@ifundefined{pdfbookmark}{\def\pdfbookmark[#1]#2#3{}}{}
\makeatother

\theoremstyle{plain}
\newtheorem*{con}{Conjecture}

\DeclareSymbolFont{AMSb}{U}{msb}{m}{n}
\DeclareMathSymbol{\N}{\mathbin}{AMSb}{"4E}
\DeclareMathSymbol{\Z}{\mathbin}{AMSb}{"5A}
\DeclareMathSymbol{\R}{\mathbin}{AMSb}{"52}
\DeclareMathSymbol{\Q}{\mathbin}{AMSb}{"51}
\DeclareMathSymbol{\I}{\mathbin}{AMSb}{"49}
\DeclareMathSymbol{\C}{\mathbin}{AMSb}{"43}
\DeclareMathSymbol{\F}{\mathbin}{AMSb}{"46}
\DeclareMathSymbol{\E}{\mathbin}{AMSb}{"45}
\DeclareMathSymbol{\K}{\mathbin}{AMSb}{"4B}
\DeclareMathSymbol{\LL}{\mathbin}{AMSb}{"4C}

\def\coloneq{\colonequals}
\def\eqcolon{\equalscolon}

\def\ket#1{{|#1\rangle}}
\def\bra#1{{\langle#1|}}
\def\ketbra#1{{|#1\rangle\langle#1|}}
\def\braket#1#2{{\langle#1|#2\rangle}}

\def\spf#1#2{{\langle#1,#2\rangle}}
\def\md#1{{\:(\operatorname{mod}\,#1)}}

\def\tr{{\operatorname{tr}}}

\def\d{\mathrm{d}}
\def\U{{\operatorname{U}}}

\def\PC{{\operatorname{PC}}}
\def\PEC{{\operatorname{PEC}}}
\def\ECl{{\operatorname{EC}}}
\def\Cl{{\operatorname{C}}}
\def\cl{{\operatorname{c}}}
\def\ecl{{\operatorname{ec}}}

\def\SLtwo{{\operatorname{SL}_2}}
\def\ESLtwo{{\operatorname{ESL}_2}}
\def\H{{\operatorname{H}}}
\def\I{{\operatorname{I}}}

\def\PCd{{\C P^{d-1}}}
\def\db{{\bar{d}}}
\def\Fz{{F_z}}
\def\Sz{{Z}}
\def\Fa{{F_a}}
\def\Fb{{F_b}}
\def\Fc{{F_c}}
\def\Fd{{F_d}}
\def\Fe{{F_e}}
\def\Fep{{F_{e'}}}
\def\orb{{\operatorname{orb}}}
\def\stab{{\operatorname{S}}}

\def\mr#1#2{\multirow{#1}{*}{#2}}
\def\mat#1#2#3#4{\bigl(\begin{smallmatrix}{#1}&{#2}\\{#3}&{#4}\end{smallmatrix}\bigr)}

\def\matcv#1#2#3#4#5#6{\bigl(\begin{smallmatrix}{#1}&{#2}\\{#3}&{#4}\end{smallmatrix}\big|\begin{smallmatrix}{#5}\\{#6}\end{smallmatrix}\bigr)}

\def\ce#1#2{{({#1}\mspace{1mu}|\mspace{1mu}{#2})}}
\def\cle#1#2{{[{#1}\mspace{1mu}|\mspace{1mu}{#2}]}}

\begin{document}

\title{SICs: Extending the list of solutions}

\author{A. J. Scott}
\email{dr.andrew.scott@gmail.com}

\begin{abstract}
 Zauner's conjecture asserts that $d^2$ equiangular lines exist in all $d$ complex dimensions.  In quantum theory, the $d^2$  lines are dubbed a SIC, as they define a favoured standard informationally complete quantum measurement called a SIC-POVM.
This note supplements A.~J.~Scott and M.~Grassl [J.\ Math.\ Phys.\ {\bf 51} (2010), 042203] by extending the list of published numerical solutions. We provide a putative complete list of Weyl-Heisenberg covariant SICs with the known symmetries in dimensions $d\leq 90$, a single solution with Zauner's symmetry for every $d\leq 121$ and solutions with higher symmetry for $d=124,143,147,168,172,195,199,228,259$ and $323$.
\end{abstract}

\maketitle

A computer study of Zauner's conjecture~\cite{Zauner99} (SIC-POVM conjecture~\cite{Renes04}) was reported by the author with M.~Grassl in ref.~\cite{Scott10}. In that article we presented a putative complete list of Weyl-Heisenberg covariant SICs in dimensions $d\leq 50$. There are now compelling reasons~\cite{Appleby13,AFMY16,AFMY17,Bengtsson16} to extend this list as far as possible, and this note sets out to do this. Background material is laid out in appendix~\ref{sec:sic}, including all notation and definitions used in the following. 

Our computer search has so far discovered SIC fiducial vectors in all dimensions $d\leq 121$ and we are confident that the list is complete for $d\leq 50$, where the search reported in ref.~\cite{Scott10} exhausted all of $\C^d$. For $d>50$, we continue the search by narrowing its scope to eigenspaces of the unitary matrices corresponding to $\Fz$ (Zauner's matrix) and $\Fa$ (defined below). Every known Weyl-Heisenberg covariant SIC in dimensions $d>3$ can be generated by a fiducial vector from one of these eigenspaces (up to extended Clifford transformations) and, within the eigenspaces, it is unlikely that solutions have been missed for $d\leq 90$ (the dimensions searched so far). Our list is therefore believed complete in these dimensions for Weyl-Heisenberg covariant SICs with the known principal symmetries, $\Fz$ and $\Fa$. Moreover, the lack of evidence from  ref.~\cite{Scott10} for other symmetries, besides those arising in conjunction with $\Fz$ and $\Fa$, indicates that it could be complete for all Weyl-Heisenberg covariant SICs.  

In higher dimensions, searching an eigenspace of Zauner's matrix discovers a solution for all $d\leq 121$\footnote{Now extended to all $d\leq 151$ by C. A. Fuchs, M. C. Hoang and B. C. Stacey using our code on the Chimera
supercomputer.}; searching eigenspaces (or coneigenvectors) of higher order symmetry matrices discovers solutions in dimensions $d=124,143,147,168,172,195,199,228,259$ and $323$. The solutions in dimensions $d> 50$ are available with 150 digits of precision in the article source files (ref.~\cite{Scott10} contains solutions for $d\leq 50$) where labelling generally follows the order that they are found. Appendix~\ref{sec:num} outlines the numerical approach to our computer search.

Symmetry structures in the new solutions can be observed. Each fiducial vector solution $\ket{\phi}\in\C^d$ that generates a Weyl-Heisenberg SIC also generates an entire orbit of related fiducial vectors under the action of the extended Clifford group:
\begin{align}
\orb(\phi) &\coloneq \{U\ketbra{\phi}U^{\dag}\}_{U\in\ECl(d)}
\end{align}
(see appendix~\ref{sec:sic} or ref.~\cite{Scott10}). In the table we list the known unique fiducial vector orbits in each dimension. Note that the length of each orbit will be $|\PEC(d)|$ unless there is a symmetry present in $\phi$, described by its stabiliser 
\begin{align}
\stab(\phi) &\coloneq \{[U]\in\PEC(d):|\bra{\phi}U\ket{\phi}|=1\}\:,
\end{align}
where $[U]\coloneq \{e^{i\xi}U\}_{\xi\in\R}$. We then have $|\orb(\phi)|=|\PEC(d)|/|\stab(\phi)|$. 

The stabiliser of each orbit is also given. The numerical initial fiducial vector
$\phi$ and its stabiliser $\stab(\phi)$ could always be chosen in a
way that the stabiliser elements take the form $\cle{F}{0}$ (where $\cle{F}{p}\coloneq [E_{\ce{F}{p}}]$ as in eq.~\ref{eq:Eclassdefn}) and, therefore, only the
matrices $F$ are quoted. Note that when $d$ is even, however,
  the subgroup of $\ESLtwo(\Z_{2d})$ that defines
  $\stab(\phi)\leq\PEC(d)$ will generally have an order that is a multiple of
  $|\stab(\phi)|$. 

In dimensions $d>3$ each stabiliser is a cyclic group of order a multiple of 3 and the vast majority, up to group conjugacy, have the symmetry described by Zauner's order-3 unitary $\cle{\Fz}{0}$, where
\begin{align}
\Fz &\coloneq \begin{pmatrix} 0 & d-1 \\ d+1 & d-1 \end{pmatrix}.
\end{align}
Exceptions occur in dimensions $d=9k+3=12(b)$, $21(e)$, $30(d)$, $39(g$--$j)$, $48(e,g)$, $57(g,h)$, $66(a$--$c)$, $75(m$--$o)$, $84(g$--$j)$, $93,\dots$, in which case solutions stabilised by the order-3 unitary $\cle{\Fa}{0}$ exist, as indicated in parentheses, where
\begin{align}
\Fa &\coloneq \begin{pmatrix} 1 & d+3 \\ d+3k & d-2 \end{pmatrix}.
\end{align}

\noindent We can find additional symmetries present in the following dimensions:
\begin{enumerate}

\item $d=k^2-1=8(b)$, $15(d)$, $24(c)$, $35(i,j)$, $48(f)$, $63(b,c)$, $80(i)$, $99$, $120$, $143(a)$, $168(a)$, $195(a)$, $224,\dots$ : order-2 unitary $\cle{\Fb}{0}$, where 
\begin{align}\label{eq:symmetryb}
\Fb &\coloneq \begin{pmatrix} -k & d \\ d & d-k \end{pmatrix}
\end{align}

\item $d=(3k\pm 1)^2+3=4(a)$, $7(b)$, $19(d,e)$, $28(c)$, $52(d)$, $67(a,b)$, $103$, $124(a)$, $172(a)$, $199$, $259(a)$, $292,\dots$ : order-2 anti-unitary $\cle{\Fc}{0}$, where 
\begin{align}
\Fc &\coloneq \begin{pmatrix} \kappa & d-2\kappa \\ d+2\kappa & d-\kappa \end{pmatrix},\quad \kappa=3k^2\pm k+1
\end{align}

\item $d=(k+3)k^2-1=19(e)$, $53(j)$, $111$, $199(a)$, $323(a)$, $489,\dots$ : order-9 unitary $\cle{\Fd}{0}$, where 
\begin{align}
\Fd &\coloneq \begin{pmatrix} 0 & 1 \\ -1 & -(k+3)k \end{pmatrix}
\end{align}
It is found that ${\Fd}^3 \sim \Fz$ for $d=19,53$ but ${\Fd}^3 \sim \Fa$ for $d=111$, where $\sim$ denotes group conjugacy.

\item $d=9k^2+3=12(b)$, $39(i,j)$, $84(i,j)$, $147(a)$, $228(a)$, $327,\dots$ : order-6 anti-unitary $\cle{\Fe}{0}$, where 
\begin{align}
\Fe &\coloneq \begin{pmatrix} 0 & 1 \\ 1 & d+3k \end{pmatrix}
\end{align}
Note that ${\Fe}^2 \sim \Fa$ since $G\Fa={\Fe}^2G$ for $G = \mat{k+1}{1}{k}{1}$. We can include $\Fc$ in this series and generalise to all dimensions $d=l^2+3=4(a)$, $7(b)$, $12(b)$, $19(d,e)$, $28(c)$, $39(i,j),\dots$, by instead setting
\begin{align}
\Fep &\coloneq \begin{cases}
    \begin{pmatrix} 0 & 1 \\ 1 & d-l \end{pmatrix},& \text{if } 3\,|\,l+1\\[1.5em]
    \begin{pmatrix} 0 & 1 \\ 1 & d+l \end{pmatrix},& \text{otherwise}
\end{cases}
\end{align}
Then $\Fep=\Fe$ when $3|l$ and $\Fep\sim\Fc\Fz$ otherwise, since $G{\Fc\Fz}={\Fep}G$ for $G = \mat{\kappa}{\kappa-1}{\kappa+1}{\kappa}$. 
In odd dimensions, i.e.\ when $l=2m$, we find that ${\Fep}^3 \sim J$ since $GJ={\Fep}^3G$ for 
\begin{align}
G &= \begin{cases}
    \begin{pmatrix} m(2m-1)+2 & -m(2m-1)-1 \\ -1 & 1 \end{pmatrix},& \text{if } 3\,|\,2m+1\\[1.5em]
    \begin{pmatrix} m(2m+1)+2 & -m(2m+1)-1 \\ -1 & 1 \end{pmatrix},& \text{otherwise}
\end{cases}
\end{align}
Therefore real fiducial vectors are expected to exist for all dimensions $d=4m^2+3$.

\end{enumerate}

\noindent Gaps in the current classification are marked by ?'s, which either indicate an unexplored dimension (to be filled in at later dates) or note the likely presence of an unknown general symmetry.

\bigskip
             
\begin{longtable}[c]{|c|c|c|c|c|c|}\hline
\mr{3}{$d$}&\multicolumn{5}{|c|}{$\PEC(d)$ orbits} \\\cline{2-6}
           & \mr{2}{\#} & \multicolumn{3}{|c|}{stabiliser} & \mr{2}{labels} \\\cline{3-5}
           &            & $|\stab|$ & $\stab$ & notes & \\\hline\hline
\endfirsthead
\hline
\mr{3}{$d$}&\multicolumn{5}{|c|}{$\PEC(d)$ orbits} \\\cline{2-6}
           & \mr{2}{\#} & \multicolumn{3}{|c|}{stabiliser} & \mr{2}{labels} \\\cline{3-5}
           &            & $|\stab|$ & $\stab$ & notes & \\\hline\hline
\endhead
\endfoot
\hline
\endlastfoot
2          & 1      & 6      & $\langle \mat{0}{-1}{-1}{0},\Fz \rangle$                                                                &  & $a$          \\\hline
\mr{3}{3}  &$\infty$& 6      & $\langle \mat{0}{-1}{-1}{0},\Fz \rangle$                                                                &  & $a$          \\           
           & 1      & 12     & $\langle \mat{0}{-1}{-1}{0},-\Fz \rangle$                                                               &  & $b$          \\           
           & 1      & 48     & $\ESLtwo(\Z_3)$                                                                                         &  & $c$          \\\hline
4          & 1      & 6      & $\langle \Fc\Fz \rangle = \langle \Fc \rangle\langle \Fz \rangle$                                       &  & $a$          \\\hline
5          & 1      & 3      & $\langle \Fz \rangle$                                                                                   &  & $a$          \\\hline
6          & 1      & 3      & $\langle \Fz \rangle$                                                                                   &  & $a$          \\\hline
\mr{2}{7}  & 1      & 3      & $\langle \Fz \rangle$                                                                                   &  & $a$          \\
           & 1      & 6      & $\langle \Fc\Fz \rangle = \langle \Fc \rangle\langle \Fz \rangle$                                       &  & $b$          \\\hline
\mr{2}{8}  & 1      & 3      & $\langle \Fz \rangle$                                                                                   &  & $a$          \\
           & 1      & 12     & $\langle \mat{6}{11}{5}{1} \rangle = \langle \Fz \rangle  \langle \mat{3}{6}{10}{9} \rangle\ni\Fb$      & ? & $b$          \\\hline
9          & 2      & 3      & $\langle \Fz \rangle$                                                                                   &  & $a,b$        \\\hline
10         & 1      & 3      & $\langle \Fz \rangle$                                                                                   &  & $a$          \\\hline
11         & 3      & 3      & $\langle \Fz \rangle$                                                                                   &  & $a$--$c$     \\\hline
\mr{2}{12} & 1      & 3      & $\langle \Fz \rangle$                                                                                   &  & $a$          \\
           & 1      & 6      & $\langle \mat{0}{17}{17}{15} \rangle\ni\Fa$                                                             & $\mat{0}{17}{17}{15} \sim \Fe$ & $b$          \\\hline
13         & 2      & 3      & $\langle \Fz \rangle$                                                                                   &  & $a,b$        \\\hline
14         & 2      & 3      & $\langle \Fz \rangle$                                                                                   &  & $a,b$        \\\hline
\mr{2}{15} & 3      & 3      & $\langle \Fz \rangle$                                                                                   &  & $a$--$c$     \\
           & 1      & 6      & $\langle \Fb\Fz \rangle = \langle\Fb\rangle\langle\Fz\rangle$                                           &  & $d$          \\\hline
16         & 2      & 3      & $\langle \Fz \rangle$                                                                                   &  & $a,b$        \\\hline
17         & 3      & 3      & $\langle \Fz \rangle$                                                                                   &  & $a$--$c$     \\\hline
18         & 2      & 3      & $\langle \Fz \rangle$                                                                                   &  & $a,b$        \\\hline
\mr{3}{19} & 3      & 3      & $\langle \Fz \rangle$                                                                                   &  & $a$--$c$     \\
           & 1      & 6      & $\langle \Fc\Fz \rangle = \langle \Fc \rangle  \langle \Fz \rangle$                                     &  & $d$          \\
           & 1      & 18     & $\langle \mat{3}{12}{7}{15} \rangle = \langle \Fc \rangle  \langle \mat{7}{14}{5}{2} \rangle \ni \Fz$   & $\mat{7}{14}{5}{2} \sim \Fd$  & $e$          \\\hline
20         & 2      & 3      & $\langle \Fz \rangle$                                                                                   &  & $a,b$        \\\hline
\mr{2}{21} & 4      & 3      & $\langle \Fz \rangle$                                                                                   &  & $a$--$d$     \\
           & 1      & 3      & $\langle \Fa \rangle$                                                                                   &  & $e$          \\\hline
22         & 1      & 3      & $\langle \Fz \rangle$                                                                                   &  & $a$          \\\hline
23         & 6      & 3      & $\langle \Fz \rangle$                                                                                   &  & $a$--$f$     \\\hline
\mr{2}{24} & 2      & 3      & $\langle \Fz \rangle$                                                                                   &  & $a,b$        \\
           & 1      & 6      & $\langle \Fb\Fz \rangle = \langle \Fb \rangle  \langle \Fz \rangle$                                     &  & $c$          \\\hline
25         & 2      & 3      & $\langle \Fz \rangle$                                                                                   &  & $a,b$        \\\hline
26         & 4      & 3      & $\langle \Fz \rangle$                                                                                   &  & $a$--$d$     \\\hline
27         & 6      & 3      & $\langle \Fz \rangle$                                                                                   &  & $a$--$f$     \\\hline
\mr{2}{28} & 2      & 3      & $\langle \Fz \rangle$                                                                                   &  & $a,b$        \\
           & 1      & 6      & $\langle \Fc\Fz \rangle = \langle \Fc \rangle\langle \Fz \rangle$                                       &  & $c$          \\\hline
29         & 4      & 3      & $\langle \Fz \rangle$                                                                                   &  & $a$--$d$     \\\hline
\mr{2}{30} & 3      & 3      & $\langle \Fz \rangle$                                                                                   &  & $a$--$c$     \\ 
           & 1      & 3      & $\langle \Fa \rangle$                                                                                   &  & $d$          \\\hline
31         & 7      & 3      & $\langle \Fz \rangle$                                                                                   &  & $a$--$g$     \\\hline
32         & 2      & 3      & $\langle \Fz \rangle$                                                                                   &  & $a,b$        \\\hline
33         & 4      & 3      & $\langle \Fz \rangle$                                                                                   &  & $a$--$d$     \\\hline
34         & 2      & 3      & $\langle \Fz \rangle$                                                                                   &  & $a,b$        \\\hline
\mr{3}{35} & 8      & 3      & $\langle \Fz \rangle$                                                                                   &  & $a$--$h$     \\ 
           & 1      & 6      & $\langle \Fb\Fz \rangle = \langle \Fb \rangle  \langle \Fz \rangle$                                     &  & $i$          \\
           & 1      & 12     & $\langle \mat{15}{3}{32}{18} \rangle = \langle \Fz \rangle  \langle \mat{3}{15}{20}{18} \rangle \ni \Fb$& ? & $j$          \\\hline
36         & 4      & 3      & $\langle \Fz \rangle$                                                                                   &  & $a$--$d$     \\\hline
37         & 4      & 3      & $\langle \Fz \rangle$                                                                                   &  & $a$--$d$     \\\hline
38         & 4      & 3      & $\langle \Fz \rangle$                                                                                   &  & $a$--$d$     \\\hline
\mr{3}{39} & 6      & 3      & $\langle \Fz \rangle$                                                                                   &  & $a$--$f$     \\
           & 2      & 3      & $\langle \Fa \rangle$                                                                                   &  & $g,h$        \\
           & 2      & 6      & $\langle \mat{0}{7}{28}{6} \rangle\ni\Fa$                                                               & $\mat{0}{7}{28}{6} \sim \Fe$ & $i,j$        \\\hline
40         & 2      & 3      & $\langle \Fz \rangle$                                                                                   &  & $a,b$        \\\hline
41         & 8      & 3      & $\langle \Fz \rangle$                                                                                   &  & $a$--$h$     \\\hline
42         & 4      & 3      & $\langle \Fz \rangle$                                                                                   &  & $a$--$d$     \\\hline\pagebreak
43         & 6      & 3      & $\langle \Fz \rangle$                                                                                   &  & $a$--$f$     \\\hline
44         & 6      & 3      & $\langle \Fz \rangle$                                                                                   &  & $a$--$f$     \\\hline
45         & 4      & 3      & $\langle \Fz \rangle$                                                                                   &  & $a$--$d$     \\\hline
46         & 3      & 3      & $\langle \Fz \rangle$                                                                                   &  & $a$--$c$     \\\hline
47         & 8      & 3      & $\langle \Fz \rangle$                                                                                   &  & $a$--$h$     \\\hline
\mr{4}{48} & 4      & 3      & $\langle \Fz \rangle$                                                                                   &  & $a$--$d$     \\
           & 1      & 3      & $\langle \Fa \rangle$                                                                                   &  & $e$          \\
           & 1      & 6      & $\langle \Fb\Fz \rangle = \langle \Fb \rangle \langle \Fz \rangle$                                      &  & $f$          \\
           & 1      & 24     & $\langle \mat{4}{37}{25}{63} \rangle\ni\Fa,\Fb$                                                         & ? & $g$          \\\hline
49         & 7      & 3      & $\langle \Fz \rangle$                                                                                   &  & $a$--$g$     \\\hline
50         & 2      & 3      & $\langle \Fz \rangle$                                                                                   &  & $a,b$        \\\hline
51         & 14     & 3      & $\langle \Fz \rangle$                                                                                   &  & $a$--$n$     \\\hline
\mr{2}{52} & 3      & 3      & $\langle \Fz \rangle$                                                                                   &  & $a$--$c$     \\
           & 1      & 6      & $\langle \Fc\Fz \rangle = \langle \Fc \rangle\langle \Fz \rangle$                                       &  & $d$          \\\hline
\mr{2}{53} & 9      & 3      & $\langle \Fz \rangle$                                                                                   &  & $a$--$i$     \\
           & 1      & 9      & $\langle \mat{7}{21}{32}{28} \rangle \ni \Fz$                                                           & $\mat{7}{21}{32}{28} \sim \Fd$ & $j$          \\\hline
54         & 4      & 3      & $\langle \Fz \rangle$                                                                                   &  & $a$--$d$     \\\hline
55         & 6      & 3      & $\langle \Fz \rangle$                                                                                   &  & $a$--$f$     \\\hline
56         & 6      & 3      & $\langle \Fz \rangle$                                                                                   &  & $a$--$f$     \\\hline
\mr{2}{57} & 6      & 3      & $\langle \Fz \rangle$                                                                                   &  & $a$--$f$     \\
           & 2      & 3      & $\langle \Fa \rangle$                                                                                   &  & $g,h$        \\\hline
58         & 4      & 3      & $\langle \Fz \rangle$                                                                                   &  & $a$--$d$     \\\hline
59         & 12     & 3      & $\langle \Fz \rangle$                                                                                   &  & $a$--$l$     \\\hline
60         & 4      & 3      & $\langle \Fz \rangle$                                                                                   &  & $a$--$d$     \\\hline
61         & 6      & 3      & $\langle \Fz \rangle$                                                                                   &  & $a$--$f$     \\\hline
62         & 5      & 3      & $\langle \Fz \rangle$                                                                                   &  & $a$--$e$     \\\hline
\mr{2}{63} & 14     & 3      & $\langle \Fz \rangle$                                                                                   &  & $a,d$--$p$   \\
           & 2      & 6      & $\langle \Fb\Fz \rangle = \langle \Fb \rangle  \langle \Fz \rangle$                                     &  & $b,c$        \\\hline
64         & 4      & 3      & $\langle \Fz \rangle$                                                                                   &  & $a$--$d$     \\\hline
65         & 8      & 3      & $\langle \Fz \rangle$                                                                                   &  & $a$--$h$     \\\hline
\mr{2}{66} & 6      & 3      & $\langle \Fz \rangle$                                                                                   &  & $d$--$i$     \\
           & 3      & 3      & $\langle \Fa \rangle$                                                                                   &  & $a$--$c$     \\\hline
\mr{2}{67} & 7      & 3      & $\langle \Fz \rangle$                                                                                   &  & $c$--$i$     \\
           & 2      & 6      & $\langle \Fc\Fz \rangle = \langle \Fc \rangle\langle \Fz \rangle$                                       &  & $a,b$     \\\hline
68         & 4      & 3      & $\langle \Fz \rangle$                                                                                   &  & $a$--$d$     \\\hline
69         & 8      & 3      & $\langle \Fz \rangle$                                                                                   &  & $a$--$h$     \\\hline
70         & 5      & 3      & $\langle \Fz \rangle$                                                                                   &  & $a$--$e$     \\\hline
71         & 18     & 3      & $\langle \Fz \rangle$                                                                                   &  & $a$--$r$     \\\hline
72         & 4      & 3      & $\langle \Fz \rangle$                                                                                   &  & $a$--$d$     \\\hline
73         & 4      & 3      & $\langle \Fz \rangle$                                                                                   &  & $a$--$d$     \\\hline
74         & 7      & 3      & $\langle \Fz \rangle$                                                                                   &  & $a$--$g$     \\\hline\pagebreak
\mr{2}{75} & 12     & 3      & $\langle \Fz \rangle$                                                                                   &  & $a$--$l$     \\
           & 3      & 3      & $\langle \Fa \rangle$                                                                                   &  & $m$--$o$     \\\hline
76         & 6      & 3      & $\langle \Fz \rangle$                                                                                   &  & $a$--$f$     \\\hline
77         & 8      & 3      & $\langle \Fz \rangle$                                                                                   &  & $a$--$h$     \\\hline
78         & 7      & 3      & $\langle \Fz \rangle$                                                                                   &  & $a$--$g$     \\\hline
79         & 14     & 3      & $\langle \Fz \rangle$                                                                                   &  & $a$--$n$     \\\hline
\mr{2}{80} & 8      & 3      & $\langle \Fz \rangle$                                                                                   &  & $a$--$h$     \\
           & 1      & 6      & $\langle \Fb\Fz \rangle = \langle \Fb \rangle  \langle \Fz \rangle$                                     &  & $i$          \\\hline
81         & 12     & 3      & $\langle \Fz \rangle$                                                                                   &  & $a$--$l$     \\\hline
82         & 3      & 3      & $\langle \Fz \rangle$                                                                                   &  & $a$--$c$     \\\hline
83         & 16     & 3      & $\langle \Fz \rangle$                                                                                   &  & $a$--$p$     \\\hline
\mr{3}{84} & 6      & 3      & $\langle \Fz \rangle$                                                                                   &  & $a$--$f$     \\
           & 2      & 3      & $\langle \Fa \rangle$                                                                                   &  & $g,h$     \\
           & 2      & 6      & $\langle \Fe \rangle$                                                                                   & ${\Fe}^2 \sim \Fa$ & $i,j$     \\\hline
85         & 4      & 3      & $\langle \Fz \rangle$                                                                                   &  & $a$--$d$     \\\hline
86         & 10     & 3      & $\langle \Fz \rangle$                                                                                   &  & $a$--$j$     \\\hline
87         & 12     & 3      & $\langle \Fz \rangle$                                                                                   &  & $a$--$l$     \\\hline
88         & 4      & 3      & $\langle \Fz \rangle$                                                                                   &  & $a$--$d$     \\\hline
89         & 10      & 3      & $\langle \Fz \rangle$                                                                                   &  & $a$--$j$     \\\hline
90         & 4      & 3      & $\langle \Fz \rangle$                                                                                   &  & $a$--$d$     \\\hline

91-121    & $\geq 1$ & $\geq 3$ &  $? \ni \Fz$    &  & $a$      \\\hline

99         &$\geq 3$&$\geq 6$& $? \ni \Fz,\Fb$ & & $b$--$d$   \\\hline
111        &$\geq 1$&$\geq 9$& $? \ni \Fd$     & ${\Fd}^3 \sim \Fa$ &         \\\hline
\mr{2}{120}&$\geq 1$&$\geq 6$& $? \ni \Fz,\Fb$ & &           \\
           &$\geq 1$&$\geq 6$& $? \ni \Fa,\Fb$ & &           \\\hline
124        &$\geq 1$&$\geq 6$& $? \ni \Fz,\Fc$ & & $a$        \\\hline
143        &$\geq 1$&$\geq 6$& $? \ni \Fz,\Fb$ & & $a$        \\\hline
147        &$\geq 1$&$\geq 6$& $? \ni \Fe$     & ${\Fe}^2 \sim \Fa$ & $a$        \\\hline
168        &$\geq 1$&$\geq 6$& $? \ni \Fz,\Fb$ & & $a$        \\\hline
172        &$\geq 1$&$\geq 6$& $? \ni \Fz,\Fc$ & & $a$        \\\hline
195        &$\geq 1$&$\geq 6$& $? \ni \Fz,\Fb$ & & $a$        \\\hline
199        &$\geq 1$&$\geq 9$& $? \ni \Fd$     & & $a$        \\\hline
228        &$\geq 1$&$\geq 6$& $? \ni \Fe$     & ${\Fe}^2 \sim \Fa$ & $a$        \\\hline
259        &$\geq 1$&$\geq 6$& $? \ni \Fz,\Fc$ & & $a$        \\\hline
323        &$\geq 1$&$\geq 9$& $? \ni \Fd$     & & $a$        \\

\end{longtable}

\newpage 

\begin{acknowledgments}

We thank Marcus Appleby, Ingemar Bengtsson, John Coates, Steven Flammia, Christopher Fuchs, Markus Grassl, Gary McConnell, Shayne Waldron and Jon Yard for their encouragement and insights.

\end{acknowledgments}

\appendix
\section{Background and Notation}
\label{sec:sic}

In simple terms, a SIC-POVM (or SIC) may be identified as a set of $d^2$ unit vectors in $d$ complex dimensions, written $\{\ket{x_1},\dots,\ket{x_{d^2}}\}\subset\C^d$, that are equiangular  
\begin{align}\label{eq:sic}
|\braket{x_j}{x_k}|^2 = \frac{d\delta_{j,k}+1}{d+1}\:.
\end{align}
Considering the rays in $\C^d$ upon which each outer product $\ketbra{x_k}$ projects, i.e.\ a set of $d^2$ equiangular lines through the origin of $\C^d$, it is also natural to identify a SIC with a subset of complex projective space, now written $\{x_1,\dots,x_{d^2}\}\subset\PCd$.

The most promising route towards a general construction of SICs involves translating a fiducial vector under the {\em Weyl displacement operators\/}:
\begin{align}
D_p &\coloneq \tau^{p_1p_2} V^{p_1} U^{p_2} \:,\qquad V\ket{k}=\ket{k+1\md{d}} \:,\qquad U\ket{k}=\omega^k\ket{k}\:,
\end{align}
where $p=(p_1,p_2)\in\Z^2$, $\tau=e^{\pi i(d+1)/d}$, $\omega=\tau^2=e^{2\pi i/d}$ (meaning $\tau^{d^2}=\tau^{2d}=\omega^d=1$), and we have fixed an orthonormal basis for $\C^d$: $\ket{0},\dots,\ket{d-1}$. Defining the symplectic form 
\begin{align}
\spf{p}{q} &\coloneq p_2q_1-p_1q_2\:, 
\end{align}
these operators obey the relations
\begin{align}
D_p D_q &= \tau^\spf{p}{q}D_{p+q} \label{eq:disprel1}\\ 
{D_p}^\dag &= D_{-p} \\
D_{p+dq} &= \begin{cases} D_p\:, & \text{if $d$ is odd;}\\ (-1)^\spf{p}{q}D_p\:, & \text{if $d$ is even,}  \end{cases}
\end{align}
and together generate a variant of the {\em Heisenberg group\/}:
\begin{align}
\H(d)\coloneq\{ e^{i\xi} D_p: p\in\Z^2,\xi\in\R \}\:.
\end{align}
Modulo its center, $\I(d)\coloneq\{e^{i\xi}I:\xi\in\R\}$, the Heisenberg group is simply a direct product of cyclic groups, $\H(d)/\I(d)\cong{\Z_d}^2$, where $\Z_d=\Z/d\Z=\{0,\dots,d-1\}$.

It was conjectured in ref.~\cite{Renes04} that, in every finite dimension, a SIC can be constructed as the orbit of a suitable {\em fiducial vector\/} $\ket{\phi}\in\C^d$ under the action of the displacement operators: 
\begin{align}
\ket{x_{p_1+p_2d+1}}=D_p\ket{\phi}\:.
\end{align}
The condition for equiangularity (\ref{eq:sic}) then becomes
\begin{align}\label{eq:whsic}
|\bra{\phi}D_p\ket{\phi}|^2 = \frac{d\delta_{p,0}+1}{d+1}\:.
\end{align}
To bolster this conjecture, such {\em Weyl-Heisenberg covariant SICs\/} were found with high numerical precision in all dimensions $d\leq 45$. Unbeknownst to the authors of ref.~\cite{Renes04}, however, a stronger conjecture had already been put forward by Gerhard Zauner in his doctoral dissertation~\cite{Zauner99}. Zauner claimed that, in every finite dimension, a fiducial vector for a Weyl-Heisenberg covariant SIC can be found in an eigenspace of the matrix
\begin{align}\label{eq:Zmatrix}
\bra{j}\Sz\ket{k}&\coloneq \frac{e^{i\xi}}{\sqrt{d}}\,\tau^{2jk+j^2}\:.
\end{align} 
\begin{con}[Zauner~\cite{Zauner99}]
In all finite dimensions there exists a fiducial vector for a Weyl-Heisenberg covariant SIC that is an eigenvector of $\Sz$.
\end{con}
Setting $\xi=\pi(d-1)/12$, it can be shown~\cite{Zauner99} that $\Sz$ has order 3: $\Sz^3=I$. The eigenspace with eigenvalue $e^{2\pi ik/3}$ will be labeled $\mathcal{Z}_k$ ($k=0,1,2$). Then
\begin{align}\label{eq:Zeignenspaces}
\dim \mathcal{Z}_k &= \lfloor (d+3-2k)/3\rfloor \:.
\end{align}

Under the action of conjugation, $\Sz$ defines an automorphism of the
Heisenberg group, $\Sz^{-1}\H(d)\Sz=\H(d)$, and therefore belongs to
the normaliser of $\H(d)$ in $\U(d)$,
\begin{align}
\Cl(d) &\coloneq\{U\in\U(d):U^{-1}\H(d)U=\H(d)\} \:,
\end{align}
which is called the {\em Clifford group\/} in quantum information theory, but more widely recognised as a variant of the {\em Jacobi group\/}. The significance of $\Cl(d)$ to SICs follows from eq.~(\ref{eq:whsic}): if $\ket{\phi}$ is a fiducial vector for a Weyl-Heisenberg covariant SIC, then so is $U\ket{\phi}$ for any $U\in\Cl(d)$.   

An explicit description of the Clifford group can be easily deduced in odd dimensions, which we now summarise. In general dimensions, for each symplectic matrix $F\in\SLtwo(\Z_d)$ and $p\in{\Z_d}^2$, let 
\begin{align}\label{eq:Fpnotation}
\ce{F}{p}=\left(\begin{array}{cc|c} F_{11} & F_{12} & p_1 \\ F_{21} & F_{22} & p_2 \end{array}\right)\coloneq\begin{pmatrix} F_{11} & F_{12} & p_1 \\ F_{21} & F_{22} & p_2 \\ 0 & 0 & 1 \end{pmatrix} ,
\end{align}
and define the matrix group
\begin{align}\label{eq:cstruct}
\cl(d) &\coloneq  \bigl\{\ce{F}{p}:F\in\SLtwo(\Z_d),p\in{\Z_d}^2\bigr\}\cong \SLtwo(\Z_d)\ltimes{\Z_d}^2\:,
\end{align}
which follows the multiplication rule
\begin{align}
\ce{F}{p}\ce{G}{q} &= \ce{FG}{p+Fq} \:.
\end{align}

Now assume $d$ is odd. Then for each $\ce{F}{p}\in\cl(d)$ there is a unique unitary $C_{\ce{F}{p}}\in\Cl(d)$, up to the multiplication of a phase $e^{i\xi}$, for which
\begin{align}\label{eq:metaconjugation}
C_{\ce{F}{p}}D_q{C_{\ce{F}{p}}}^\dag &= \omega^\spf{p}{Fq}D_{Fq}
\end{align}
for all $q\in\Z^2$. All Clifford operators take this action and, in fact, $C_g$ is a faithful projective unitary representation of $\cl(d)$ (i.e.\ $C_gC_h=e^{i\xi(g,h)}C_{gh}$ for some $\xi:\cl(d)\times\cl(d)\rightarrow\R$) 
that defines the group isomorphism
\begin{align}
\cl(d) &\cong \Cl(d)/\I(d)\:.
\end{align}
The Clifford group describes all automorphisms of $\H(d)$ that leave its center pointwise fixed: the inner automorphisms are just displacement operators, $D_p=C_{\ce{I}{p}}$, while the outer automorphisms are specified by the operators $C_{\ce{F}{0}}$, which are more widely recognised as {\em metaplectic operators\/}. Since eq.~(\ref{eq:metaconjugation}) defines each Clifford operator uniquely, up to a phase $e^{i\xi}$, it can be used to derive formulas, e.g., by multiplying on the right by ${D_{q}}^\dag$ and summing over $q$ to obtain  
\begin{align}\label{eq:oddCdefn}
C_{\ce{F}{p}} &= \frac{e^{i\xi}}{d\sqrt{\eta(F)}}\,D_{p}\sum_r D_{Fr}{D_{r}}^\dag\:,
\end{align}
where $\eta(F)\coloneq |\{q\in{\Z_d}^2|Fq=q\}|=|\tr\,C_{\ce{F}{0}}|^2$. To check eq.~(\ref{eq:metaconjugation}), simply replace $r$ by $r+q$ in the sum, and use $\spf{Fr}{Fq}=(\det F)\spf{r}{q}$ and eq.~(\ref{eq:disprel1}) repeatedly to obtain:
\begin{align}
C_{\ce{F}{p}} &= \omega^\spf{p}{Fq}D_{Fq}\,\frac{e^{i\xi}D_{p}}{d\sqrt{\eta(F)}}\sum_r\tau^{(\det F-1)\spf{r}{q}} D_{Fr}{D_{r}}^\dag\,{D_{q}}^\dag \\
&= \omega^\spf{p}{Fq}D_{Fq}C_{\ce{F}{p}}{D_{q}}^\dag\:,
\end{align}
since $\det F=1 \md{d}$ and $\tau^d=1$ for odd $d$. In even dimensions we would need $\det F=1 \md{2d}$, however, or we would need to generalise eq.~(\ref{eq:metaconjugation}) so that the factor $\tau^{(\det F-1)q_1q_2}$ appears on the right. The former approach was taken by Appleby~\cite{Appleby05} which we now summarise. 

In even dimensions, the metaplectic operators can introduce sign changes to eq.~(\ref{eq:metaconjugation}) unless we instead take $F\in\SLtwo(\Z_{2d})$. We also take $p\in{\Z_{2d}}^2$ for simplicity. In general dimensions, let
\begin{align}
\db &\coloneq \begin{cases} d\:, & \text{if $d$ is odd;}\\ 2d\:, & \text{if $d$ is even,}  \end{cases}
\end{align}
and identify
\begin{align}
F &= \begin{pmatrix} \alpha & \beta \\ \gamma & \delta \end{pmatrix} \in \SLtwo(\Z_{\db}) \:.
\end{align}
Now, and hereafter, for each $\ce{F}{p}\in\cl(\db)$ define
\begin{align}\label{eq:genCdefn}
C_{\ce{F}{p}} &\coloneq D_p V_F\:,
\end{align}
with
\begin{align}\label{eq:Vdefn}
\bra{j}V_F\ket{k} \coloneq \frac{1}{\sqrt{d}}\,\tau^{\beta^{-1}\left(\alpha k^2-2jk+\delta j^2\right)}\,
\end{align}
if there exists an element $\beta^{-1}\in\Z_{\db}$ with $\beta^{-1}\beta=1\md{\db}$; otherwise, we find an integer $x\in\Z_{\db}$ with the property that $(\delta+x\beta)^{-1}\in\Z_{\db}$ (whose existence is guaranteed~\cite[Lemma~3]{Appleby05}) and take  $V_F=V_{F_1}V_{F_2}$, using eq.~(\ref{eq:Vdefn}) now for $V_{F_1}$ and $V_{F_2}$, where
\begin{align}
F_1 &= \begin{pmatrix} 0 & -1 \\ 1 & x \end{pmatrix} \quad\text{and}\quad F_2 = \begin{pmatrix} \gamma+x\alpha & \delta+x\beta \\ -\alpha & -\beta \end{pmatrix}
\end{align}
fulfill the decomposition $F=F_1F_2$. In odd dimensions, eq.~(\ref{eq:genCdefn}) is equivalent to eq.~(\ref{eq:oddCdefn}). In all dimensions, now choosing $\xi=0$ in eq.~(\ref{eq:Zmatrix}), Zauner's matrix is $\Sz=C_{\ce{\Fz}{0}}$, where
\begin{align}\label{eq:Fz}
\Fz &\coloneq \begin{pmatrix} 0 & d-1 \\ d+1 & d-1 \end{pmatrix}.
\end{align}

With these definitions, Appleby~\cite[Theorem~1]{Appleby05} showed that the map $g\mapsto C_g$ defines a group isomorphism 
\begin{align}\label{eq:Cstruct}
\cl(\db)/\ker(C) \cong\Cl(d)/\I(d) \eqcolon \PC(d)\:, 
\end{align}
obeying eq.~(\ref{eq:metaconjugation}), where
\begin{align}
\ker(C) &= \begin{cases} \bigl\{ \matcv{1}{0}{0}{1}{0}{0} \bigr\}\:, & \text{if $d$ is odd;}\\ \bigl\langle\matcv{1+d}{0}{0}{1+d}{0}{0},\matcv{1}{d}{0}{1}{d/2}{0},\matcv{1}{0}{d}{1}{0}{d/2}\bigr\rangle\:, & \text{if $d$ is even.}  \end{cases}
\end{align}
Combining eqs.~(\ref{eq:cstruct}) and (\ref{eq:Cstruct}) we can deduce the size of the Clifford group. It is known that
\begin{align}
|\SLtwo(\Z_d)|=d^3\prod_{p\mid d}\bigl(1-p^{-2}\bigr)\:,
\end{align}
where the product is over all primes $p$ dividing $d$, which means $|\cl(\db)|=32|\cl(d)|$ for even $d$. But since $|{\ker(C)}|=32$ in even dimensions, we can conclude that
\begin{align}
|\PC(d)|=|\cl(d)|=|\SLtwo(\Z_{d})||{\Z_{d}}^2|=d^5\prod_{p\mid d}\bigl(1-p^{-2}\bigr)
\end{align}
in all dimensions.

Finally, note that there is a complex-conjugation symmetry apparent in eq.~(\ref{eq:whsic}). Let $\hat{J}=\hat{J}^\dag$ be the anti-unitary operator with action $\hat{J}\sum_k c_k\ket{k}=\sum_k {c_k}^*\ket{k}$ for a vector rewritten in our standard basis. Then, 
\begin{align}
\hat{J} D_p \hat{J}^\dag &= D_{Jp}
\end{align}
for all $p\in\Z^2$, where
\begin{align}
J &\coloneq \begin{pmatrix} 1 & 0 \\ 0 & -1 \end{pmatrix}.
\end{align}
It thus follows from eq.~(\ref{eq:whsic}) that $\hat{J}\ket{\phi}$ is a fiducial vector for a Weyl-Heisenberg covariant SIC whenever $\ket{\phi}$ is. To analyze this additional symmetry, define the matrix group
\begin{align}\label{eq:ecstruct}
\ecl(d) &\coloneq  \bigl\{\ce{F}{p}:F\in\ESLtwo(\Z_d),p\in{\Z_d}^2\bigr\}\cong \ESLtwo(\Z_d)\ltimes{\Z_d}^2\:,
\end{align}
where $\ESLtwo(\Z_d)\coloneq\bigl\{F\in{\operatorname{Mat}_{2,2}}(\Z_d):\det F=\pm 1\md{d}\bigr\}=\SLtwo(\Z_d)\cup J\,\SLtwo(\Z_d)$ is the union of all symplectic and anti-symplectic matrices. The {\em extended Clifford group\/} is then defined as the group of all unitary and anti-unitary operators that normalise $\H(d)$, i.e., the disjoint union 
\begin{align}\label{eq:ECdefn}
\ECl(d) &\coloneq \Cl(d)\cup \hat{J}\,\Cl(d) \:.
\end{align}
Appleby~\cite[Theorem~2]{Appleby05} showed that 
\begin{align}\label{eq:ECstruct}
\ecl(\db)/\ker(E) \cong\ECl(d)/\I(d) \eqcolon \PEC(d)\:, 
\end{align} 
through the map $E:\ecl(\db)\rightarrow\ECl(d)$, where
\begin{align}\label{eq:genEdefn}
E_{\ce{F}{p}} &\coloneq  \begin{cases} C_{\ce{F}{p}}\:, & \text{if $\ce{F}{p}\in\cl(\db)$;}\\ \hat{J}C_{\ce{JF}{Jp}}\:, & \text{otherwise.}  \end{cases}
\end{align}
From eq.~(\ref{eq:ECdefn}), we know that $\ker(E)=\ker(C)$ and thus have $|\PEC(d)|=2|\PC(d)|$. 

Lastly, in eqs.~(\ref{eq:Cstruct}) and (\ref{eq:ECstruct}) we have
defined projective versions of the Clifford and extended Clifford
groups, respectively. The notation $[U]\coloneq
\{e^{i\xi}U\}_{\xi\in\R}$ is used to
denote members of these groups, which are equivalence classes of
unitary and anti-unitary matrices that differ only by a factor of unit modulus.  We also use the shorthand notation
\begin{align}\label{eq:Eclassdefn}
\cle{F}{p} &= \left[\begin{array}{cc|c} F_{11} & F_{12} & p_1 \\ F_{21} & F_{22} & p_2 \end{array}\right] \coloneq [E_{\ce{F}{p}}]\in\PEC(d) \:.
\end{align}

See ref.~\cite{Scott10} for more.

\section{Numerical Methods}
\label{sec:num}

Another characterisation of SICs comes from design theory. A finite set $\mathscr{D}\subset\PCd$ is called a {\em complex projective $t$-design\/} if
\begin{align}\label{eq:tdesign}
\frac{1}{|\mathscr{D}|}\sum_{x\in\mathscr{D}}\ketbra{x}^{\otimes t}&=\mathop{\int}_{\PCd}\d\mu(x)\,\ketbra{x}^{\otimes t}\:,
\end{align}
where $\mu$ is the Haar measure. In these terms, SICs are precisely equivalent to {\em tight\/} complex projective 2-designs, which are 2-designs that 
meet the absolute bound on their size: $|\mathscr{D}|\geq d^2$. All 2-designs with $|\mathscr{D}|=d^2$ are necessarily sets of equiangular lines (SICs) 
and these are the only 2-designs with this structure.

Our numerical approach makes use of the {\em Welch bound\/}: for any finite $\mathscr{S}\subset\PCd$, and any positive integer $t$,
\begin{align}\label{eq:welchbound}
\frac{1}{|\mathscr{S}|^2}\sum_{x,y\in\mathscr{S}}\,|\braket{x}{y}|^{2t} &\geq \tbinom{d+t-1}{t}^{-1}\:,
\end{align}
with equality if and only if $\mathscr{S}$ is a $t$-design. Setting $\ket{x_{p_1+p_2d+1}}=D_p\ket{\phi}$, $|\mathscr{S}|=d^2$ and $t=2$, we can 
translate eq.~(\ref{eq:welchbound}) to our case: {\it for any\/ $\ket{\phi}\in\C^d$,
\begin{align}\label{eq:sicbound}
\frac{1}{d^3}\sum_{q,p}|\bra{\phi}{D_q}^\dag D_p\ket{\phi}|^4=\sum_{j,k}\Big|\sum_l\braket{\phi}{j+l}\braket{l}{\phi}\braket{\phi}{k+l}\braket{j+k+l}{\phi}\Big|^2 &\geq \frac{2}{d+1}\:,
\end{align}
with equality if and only if\/ $\ket{\phi}$ is a fiducial vector for a Weyl-Heisenberg covariant SIC.\/}

We search for SIC fiducial vectors by minimising the LHS of the inequality in eq.~(\ref{eq:sicbound}), parametrised as a function of the real and 
imaginary parts of the $d$ complex numbers $\braket{j}{\phi}$, until the bound on the RHS is met. In practice, this is done by repeating a local 
search from different initial trial vectors until a solution is found. The local search uses a C++ implementation of L-BFGS. Generally, local minima are found quickly but an enormous number of trials are 
required to locate the global minima, i.e. the SICs. 

In order to exhaust the search space to arrive at a putative complete list of unique SICs, we choose initial points uniformly at random under the unitarily invariant 
Haar measure on $\PCd$ (using the Hurwitz parametrisation). Once enough solutions are found, and their precision refined (using GMP/MPFR multi-precision arithmetic), we identify unique extended-Clifford orbits by comparing each 
solution $\ket{\phi}$ to every other translated solution $E_{\ce{F}{p}}\ket{\phi'}$ with $E_{\ce{F}{p}}\in\ECl(d)$. This brute force approach to the classification, while stubborn, gives the 
symmetries (i.e.\ the stabiliser $\stab(\phi)$) essentially for free, if $\ket{\phi}$ is also compared to $E_{\ce{F}{p}}\ket{\phi}$.

Any symmetry conjectured to exist in general dimensions allows us to lower the dimension of the search space without much effort. For suppose 
$U\ket{\psi}$ = $\lambda\ket{\psi}$ for some unitary $U$ of order $n$, $U^n=I$. Then the projector onto the eigenspace with eigenvalue 
$\lambda=e^{2\pi im/n}$, where we will have $m$ integer, is 
\begin{align}
P = \frac{1}{n}\sum_{j=0}^{n-1} e^{-\frac{2\pi i j m}{n}}U^j
\end{align}
and we can search this subspace by simply replacing $\ket{\phi}$ with $\ket{\psi}=P\ket{\phi}$ in eq.~(\ref{eq:sicbound}).

There is a similar approach to anti-unitary symmetries. Suppose instead that $\hat{J}U\ket{\psi}$ = $\lambda\ket{\psi}$ for some unitary $U$ 
with $\big(\hat{J}U\big)^{2n}=\big(\hat{J}U\hat{J}U\big)^n=\big(U^*U\big)^n=I$, where $U^*$ denotes complex conjugation of matrix components 
in the standard basis. We must have $|\lambda|=1$ and may in fact assume $\lambda=1$ for coneigenvalues. Now given that the projector onto the 
eigenspace of the unitary $U^*U$ with eigenvalue 1 is
\begin{align}
Q = \frac{1}{n}\sum_{j=0}^{n-1} \big(U^*U\big)^j
\end{align}
it is easy to check that the (unnormalised) $\ket{\psi'} = U^*Q^*\ket{\phi^*}+Q\ket{\phi}$ solves our coneigenvalue problem. We can therefore 
replace $\ket{\phi}$ with $\ket{\psi}=\ket{\psi'}/\sqrt{\braket{\psi'}{\psi'}}$ in eq.~(\ref{eq:sicbound}) to search the set of coneigenvetors.

\section{Solutions}
\label{sec:sol}

Contents of the folder \texttt{../solutions} \\

\newcounter{c} 
\setcounter{c}{0}
\noindent
\foreach \d in {51,...,350}{%
  \foreach \l in {a,...,z}{%
    \IfFileExists{solutions/sicfiducial.\d.\l.150.txt}{%
      \small{\texttt{sicfiducial.\d.\l.150.txt}}%
      \addtocounter{c}{1}%
      \ifnum%
        \value{c}=3%
        \\%
        \setcounter{c}{0}%
      \else%
        \hspace{0.5cm}%
      \fi%
    }%
  }%
}%

\end{document}